\documentclass[12pt,a4paper]{article}

\usepackage{amsmath, amssymb}

\begin{document}

\title{The algebraic origin of the Doppler factor in the Li\'{e}nard-Wiechert potentials}
\author{C\u alin Galeriu}

\maketitle

\section*{Abstract}

After reviewing the algebraic derivation of the Doppler factor in the Li\'{e}nard-Wiechert potentials
of an electrically charged point particle,
we conclude that the Dirac delta function used in electrodynamics must be the one obeying the weak
definition, non-zero in an infinitesimal neighborhood, and not the one obeying the strong definition,
non-zero in a point. This conclusion emerges from our analysis of 
a) the derivation of an important Dirac delta function identity, which generates the Doppler factor,
b) the linear superposition principle implicitly used by the Green function method, and
c) the two equivalent formulations of the Schwarzschild-Tetrode-Fokker action.
As a consequence, in full agreement with our previous discussion of the geometrical origin of the Doppler factor,
we conclude that the electromagnetic interaction takes place not between points in Minkowski space,
but between corresponding infinitesimal segments along the worldlines of the particles. 

\section{Introduction}

We have recently investigated the geometrical origin of the Doppler factor in the Li\'{e}nard-Wiechert (LW)
potentials \cite{Galeriu2021a,Galeriu2021b}. 
For an extended charged particle,
the geometrical origin of the Doppler factor was well understood.
In this case, first we had to find 
the intersection of the worldtube of the charged particle with the past lightcone with
vertex at the field point,
and then we had to project this intersection onto the 3D space of the observer.
For a point charged particle,
the geometrical origin of the Doppler factor was not as obvious.
In this case it seemed that the electromagnetic interaction was taking place not
between point particles in Minkowski space, but between length elements
along the worldlines of the charged particles, with the endpoints of these 
corresponding infinitesimal segments being connected by light signals.
Since the material point particle model has failed to provide an intuitive explanation of the Doppler factor, 
we have suggested to replace the four-force
acting on a point particle with a four-force linear density acting on a
length element along the worldline of that particle.

We would have expected to reach a similar conclusion when considering 
the algebraic derivation of the Doppler factor in the LW potentials.
But no such a conclusion has been proposed yet. 
How could a geometrical treatment of a physics problem arrive at a conclusion that
was not also reached by an algebraic treatment of the same problem?
In order to elucidate this apparent paradox, we had to address some conceptual issues
afflicting the widely used Dirac delta function.

\section{The Li\'{e}nard-Wiechert potentials}

Assuming that the Lorenz gauge condition applies,
the electromagnetic four-potential $A^\alpha = (\phi, {\bf A})$ 
of a charged particle in motion is given by \cite{Jackson1999}
\begin{equation}
A^\alpha (x) = \frac{4 \pi}{c} \int G(x, x') \, J^\alpha (x') \, d^4x',
\label{eq:Green_fn_solution}
\end{equation}
where $x = (c t, {\bf x})$, $x'= (c t', {\bf x'})$, and $G(x, x')$ is the Green function. 
For a point particle with electric charge $q$, position four-vector $X^\alpha = (c T, {\bf X})$, 
four-velocity $U^\alpha = (\gamma c, \gamma {\bf V})$, and proper time $\tau$, the four-current density is given by \cite{Jackson1999}
\begin{equation}
J^\alpha(x') = \int q \, U^\alpha(\tau) \, \delta^{4D}\big( x' - X(\tau) \big) c \, d\tau.
\label{eq:current_density}
\end{equation}

The Green function must satisfy the equation
\begin{equation}
\Box_x G(x, x') = \delta^{4D}(x - x'),
\label{eq:Green_fn_equation}
\end{equation}
which is most often solved using the method of the Fourier transform. 
However, since the integrand is singular, \lq\lq Green functions that differ in their behavior are obtained by choosing
different contours of integration relative to the poles\rq\rq \cite{Jackson1999}. 
Not only the retarded and the advanced Green functions thus found are solutions 
of (\ref{eq:Green_fn_equation}), but also any
linear combination whose coefficients add up to one.
All these Green functions share the property that
\lq\lq the differences between two solutions of the inhomogeneous equation is a solution
of the homogeneous equation\rq\rq \cite{Thirring1958}.

We could also start by noticing that a possible solution of (\ref{eq:Green_fn_equation}) is \cite{Anderson1967, Ryder1974}
\begin{equation}
G(x, x') = \frac{1}{4 \pi} \delta\big( (x - x')^2 \big) = \frac{1}{4 \pi} \delta\big( c^2 (t - t')^2 - ({\bf x} - {\bf x'})^2 \big). 
\label{eq:symmetric_Green_fn}
\end{equation}
Next we use the Dirac delta function identity
\begin{equation}
\delta(w^2 - a^2) = \frac{1}{2 |a|} \big( \delta(w - a) + \delta(w + a) \big),
\label{eq:Dirac_delta_property_1}
\end{equation}
and, with $r = |{\bf x} - {\bf x'}|$, we rewrite the Green function (\ref{eq:symmetric_Green_fn}) as
one half of the retarded Green function (non-zero when $t'= t - r/c$)
plus one half of the advanced Green function (non-zero when $t'= t + r/c$).
\begin{equation}
G(x, x') = \frac{1}{2} \frac{\delta(c t - c t' - r)}{4 \pi r} + \frac{1}{2}\frac{\delta(c t - c t' + r)}{4 \pi r}.
\label{eq:symmetric_Green_fn_2}
\end{equation}
According to our previous discussion, a possible solution could also be
\begin{equation}
G(x, x') = f \frac{\delta(c t - c t' - r)}{4 \pi r} + ( 1 - f ) \frac{\delta(c t - c t' + r)}{4 \pi r},
\label{eq:symmetric_Green_fn_3}
\end{equation}
where $f$ is a yet undetermined real number.

When Li\'{e}nard \cite{lienard} and Wiechert \cite{wiechert} have derived their electromagnetic potentials, 
they have only considered the causal solution.
From a mathematical point of view, this corresponds to setting $f = 1$ in (\ref{eq:symmetric_Green_fn_3}).
Equivalently, this also corresponds to replacing the $( 1 - f )$ contribution of the
advanced Green function with a retarded Green function of the same weight.
Why are we allowed to replace an advanced electromagnetic potential contribution with a retarded one?
The advanced potential is exactly equal to the retarded potential when the source particle is
at rest, or in motion with constant velocity, or in motion with constant acceleration (hyperbolic motion).
On top of that, by design, classical electrodynamics ignores any effect that the rate of change 
in the acceleration of the source, or of the test particle,
may have on the electromagnetic interaction (the Lorentz four-force). 
However, as if in opposition to the above statements, 
we also notice that a consistent treatment of a system of two particles in circular motion 
requires a time symmetric interaction \cite{Schild1963}.

Substitution of the causal Green function 
and of the four-current density
into (\ref{eq:Green_fn_solution}) gives
\begin{multline}
A_{(ret)}^\alpha (x) = \frac{4 \pi}{c} \int \frac{\delta(c t - c t' - r)}{4 \pi r} \, 
\int q \, U^\alpha(\tau) \, \delta^{4D}\big( x' - X(\tau) \big) c \, d\tau \, d^4x' \\
= q \int \frac{\delta(c t - c T - R)}{R} \, U^\alpha(\tau) \, d\tau
= q \int \frac{\delta(T - t + R/c)}{R} \, (1, {\bf V}/c)^\alpha \, dT,
\label{eq:Green_fn_solution_2}
\end{multline}
where $R = |{\bf x} - {\bf X}|$, since $x' = X(\tau)$ after the integration over $d^4x'$.
In the last step we have used the Dirac delta function identity
\begin{equation}
\delta(- a w) = \delta(a w) = \frac{1}{|a|} \delta(w),
\label{eq:Dirac_delta_property_2}
\end{equation}
and in the next step we use another Dirac delta function identity
\begin{equation}
\delta\big( g(w) \big) = \sum_n \frac{\delta(w - w_n)}{|g'(w_n)|},
\label{eq:Dirac_delta_property_3}
\end{equation}
where $g(w_n) = 0$ and the derivatives $g'(w_n) \neq 0$.

It is precisely at this step that the Doppler factor in the retarded LW potentials emerges. 
In our case the function $g$ is \cite{Zangwill2013}
\begin{equation}
g(T) = T - t + \frac{R(T)}{c} = T - t + \frac{\sqrt{\big({\bf x} - {\bf X}(T)\big)^2}}{c},
\label{eq:function}
\end{equation}
the retarded time $T_r$ is the unique solution for which $g(T_r) = 0$, 
\begin{equation}
T_r = t - \frac{R(T_r)}{c},
\label{eq:ret_time}
\end{equation}
the derivative $g'$ of the function $g$ is
\begin{equation}
g'(T) = 1 - \frac{{\bf V}(T) \cdot {\bf R}(T)}{c \, R(T)},
\label{eq:derivative}
\end{equation}
where ${\bf R} = {\bf x} - {\bf X}$, and the reciprocal of the Doppler factor is
\begin{equation}
g'(T_r) = 1 - \frac{{\bf V}(T_r) \cdot {\bf R}(T_r)}{c \, R(T_r)} > 0.
\label{eq:reciprocal_1}
\end{equation}

After the integration over $dT$, the retarded LW potentials of a charged point particle are obtained
\begin{equation}
(\phi_{(ret)}, {\bf A}_{(ret)}) 
= \frac{q}{R(T_r)} \frac{1}{1 - \frac{{\bf V}(T_r) \cdot {\bf R}(T_r)}{c \, R(T_r)}} \big( 1, {\bf V}(T_r)/c \big).
\label{eq:LWpotentials}
\end{equation}

We also notice that, from (\ref{eq:ret_time}), we can express $t$ as a function of $T_r$. 
\begin{equation}
t(T_r) = T_r + \frac{R(T_r)}{c}.
\label{eq:field_point_time}
\end{equation}
It follows that $t'(T_r)$, the derivative of $t$
with respect to $T_r$, is exactly equal to $g'(T_r)$, the reciprocal of the Doppler factor. 

Substitution of the acausal Green function 
and of the four-current density
into (\ref{eq:Green_fn_solution}) gives
\begin{multline}
A_{(adv)}^\alpha (x) = \frac{4 \pi}{c} \int \frac{\delta(c t - c t' + r)}{4 \pi r} \, 
\int q \, U^\alpha(\tau) \, \delta^{4D}\big( x' - X(\tau) \big) c \, d\tau \, d^4x' \\
= q \int \frac{\delta(c t - c T + R)}{R} \, U^\alpha(\tau) \, d\tau
= q \int \frac{\delta(T - t - R/c)}{R} \, (1, {\bf V}/c)^\alpha \, dT,
\label{eq:Green_fn_solution_2_adv}
\end{multline}
and, after a similar calculation, the advanced LW potentials of a charged point particle are obtained
\begin{equation}
(\phi_{(adv)}, {\bf A}_{(adv)}) 
= \frac{q}{R(T_a)} \frac{1}{1 + \frac{{\bf V}(T_a) \cdot {\bf R}(T_a)}{c \, R(T_a)}} \big( 1, {\bf V}(T_a)/c \big),
\label{eq:LWpotentials_adv}
\end{equation}
where the advanced time $T_a$ is
\begin{equation}
T_a = t + \frac{R(T_a)}{c}.
\label{eq:adv_time}
\end{equation}
Again, it follows that the derivative of $t$
with respect to $T_a$ is exactly equal to the reciprocal of the Doppler factor. 

\section{Point particle or length element?}

The material point particle model seems to be in agreement with the above derivations.
Consistent with (\ref{eq:current_density}), the electric charge density is given by
\begin{equation}
\rho({\bf x'}, t') = q \, \delta^{3D}\big( {\bf x'} - {\bf X}(t') \big),
\label{eq:charge_density}
\end{equation}
and the electric current density is given by
\begin{equation}
{\bf J}({\bf x'}, t') = q \, {\bf V}(t') \, \delta^{3D}\big( {\bf x'} - {\bf X}(t') \big).
\label{eq:3D_current_density}
\end{equation}
It appears that the charged particle is indeed a point particle, 
since the Dirac delta function is not equal to zero in just one point.
By definition \cite{Dirac1927} 
\begin{eqnarray}
\delta(w) = 0 \ \ \  \text{when} \ \ \ w \neq 0, \label{eq:Dirac_delta_a} \\
\int_{- \infty}^{\infty} \delta(w) \, dw = 1. \label{eq:Dirac_delta_b}
\end{eqnarray}

In order to gain more insight, we focus on the derivation step that has produced the Doppler factor,
which is the application of formula (\ref{eq:Dirac_delta_property_3}). 
How can a Dirac delta function, which is non-zero in just one point, return the value of a derivative?
To calculate the derivative of a function in a point we need to know the value of that function in an
infinitesimal neighborhood of that point, the information from just one point is not enough.
And how was formula (\ref{eq:Dirac_delta_property_3}) derived in the first place? 
The proof involves an integration by substitution. A closer look at this method of integration
reveals the fact that the function in the integrand must be continuous, a property that the 
Dirac delta function (\ref{eq:Dirac_delta_a}) certainly does not have. In fact, the Dirac delta function must also
be a differentiable function, since from this property it follows that 
the electric charge density (\ref{eq:charge_density})
and the electric current density (\ref{eq:3D_current_density})
satisfy the continuity equation \cite{Landau4ed}. 
Dirac explicitly assumed that the delta function is differentiable, also also provided a second,
quite different description of his function. This time the Dirac delta function \lq\lq is equal to zero except when 
$x$ is very small\rq\rq \ \cite{Dirac1927}. Equation (\ref{eq:Dirac_delta_a}) is replaced by
\begin{equation}
\delta(w) = 0 \ \ \  \text{when} \ \ \ |w| > \epsilon, 
\label{eq:Dirac_delta_c}
\end{equation}
where $\epsilon$ is a positive infinitesimal number. 

We are, in fact, given two very different definitions of the same mathematical object! 
This confusing situation has been recognized by Amaku {\it et al.} \cite{Amaku2021RBEF},
who have also discussed the criterion that allows us to distinguish one type of Dirac delta function from the other.
The Dirac delta function that is non zero in just one point, corresponding to the strong definition (\ref{eq:Dirac_delta_a}),
has associated with it an integral from zero to infinity of value one 
\begin{equation}
\int_{0}^{\infty} \delta(w) \, dw = 1.
\label{eq:Dirac_strong_def}
\end{equation}
The Dirac delta function that is non zero in the infinitesimal  neighborhood of a point, 
corresponding to the weak definition (\ref{eq:Dirac_delta_c}),
has associated with it an integral from zero to infinity of value one half
\begin{equation}
\int_{0}^{\infty} \delta(w) \, dw = \frac{1}{2}.
\label{eq:Dirac_weak_def}
\end{equation}

Which one of these two definitions actually applies to the 
Dirac delta function used in our calculations of the LW potentials? 
It is the weak definition of the Dirac delta function, as it is revealed by the proof of the 
Dirac delta function identity  (\ref{eq:Dirac_delta_property_3}).
In the simple case when the function $g(w)$ has only one root $w_o$, with $g(w_o) = 0$, 
this proof consists of two steps.

Step I. For a given test function $f(w)$, the domain of integration of the integral $\int \delta(g(w)) \ f(w) \ dw$ 
is restricted to an infinitesimal domain centered on the root $w_o$, since everywhere else the delta function is zero.
Inside this infinitesimal domain the function $g(w)$ is equal to its first order Taylor series approximation.
\begin{equation}
g(w) = g(w_o) + (w - w_o) \frac{dg}{dw}\Big|_{w_o} = (w - w_o) \, g'(w_o).
\end{equation}

Step II. A change of variables gets the constant factor $|g'(w_o)|$ out of the Dirac delta function, 
a step equivalent to using the identity
\begin{equation}
\delta\big(a (w - w_o)\big) = \frac{1}{|a|} \delta(w - w_o).
\end{equation}

In Step I we have assumed that the Dirac delta function in non-zero in an infinitesimal neighborhood of a point.
In Step II we have assumed that the Dirac delta function is continuous, as required by the integration by substitution.
Only the weak definition (\ref{eq:Dirac_delta_c}) is consistent with these two assumptions.

\section{An analysis of the Green function method}

The same conclusion regarding the true nature of the Dirac delta function 
used in electrodynamics
can be reached from a different perspective, 
from an analysis of the Green function method that we have
used in order to find the electromagnetic LW potentials.
For simplicity, we only discuss the one dimensional problem.
The same conclusion holds when we use the Green function method in higher dimensions.

Consider a linear differential operator $\mathcal{L}$ and the ordinary differential equation
\begin{equation}
\mathcal{L} y(x) = f(x).
\label{eq:ODE}
\end{equation}
The Green function satisfies the equation
\begin{equation}
\mathcal{L} G(x, x') = \delta(x - x').
\label{eq:Green_fn_1D}
\end{equation}
The solution of the ordinary differential equation is calculated as
\begin{equation}
y(x) = \int G(x, x') \, f(x') \, dx'.
\label{eq:ODE_solution}
\end{equation}
The solution (\ref{eq:ODE_solution}) is usually justified by showing that, 
when the operator $\mathcal{L}$ is applied on both sides
of equation (\ref{eq:ODE_solution}), the original equation (\ref{eq:ODE}) is obtained. 
This extremely short and abstract proof does not seem to be conducive to intuitive understanding.

More insight comes from a comment by Arfken {\it et al.} \cite{Arfken}, who mention that
\lq\lq The fact that we can determine $\psi$ everywhere by an integration is a consequence
of the fact that our differential equation is linear, so each element of the source contributes additively.\rq\rq

Inspired by this comment, we generate an infinite set of points $x_k = 2 k \epsilon$, 
where $k$ spans the set of
integer numbers, and we divide the $x$-axis into an infinite set of 
infinitesimal intervals $[x_k - \epsilon, x_k + \epsilon)$. The source term is now written as
\begin{equation}
f(x) = \sum_k f_k(x),
\label{eq:source_term_sum}
\end{equation}
where
\begin{equation}
f_k(x) = 
\begin{cases}
0, &x < x_k - \epsilon \\
f(x), &x_k - \epsilon \leq x < x_k + \epsilon \\
0, &x_k + \epsilon \leq x \\
\end{cases}
\label{eq:source_term_def}
\end{equation}
The equation (\ref{eq:ODE}) splits into a set of equations
\begin{equation}
\mathcal{L} y_k(x) = f_k(x),
\label{eq:ODE_set}
\end{equation}
and, due to the linearity of $\mathcal{L}$, the total solution is obtained as a sum
\begin{equation}
y(x) = \sum_k y_k(x).
\label{eq:solution_sum}
\end{equation}

In order to show that the sum in (\ref{eq:solution_sum}) is equal to the integral in (\ref{eq:ODE_solution}), 
we start by introducing the Dirac delta function
\begin{equation}
\delta(x - x_k) = 
\begin{cases}
0, &x < x_k - \epsilon \\
\frac{1}{2 \epsilon}, &x_k - \epsilon \leq x < x_k + \epsilon \\
0, &x_k + \epsilon \leq x \\
\end{cases}
\label{eq:Dirac_delta_d}
\end{equation}

We notice that
\begin{equation}
f_k(x) = f(x) \, \delta(x - x_k) \, 2 \epsilon,
\label{eq:33}
\end{equation}
and also that, in the limit $\epsilon \to 0$,
\begin{equation}
f_k(x) = f(x_k) \, \delta(x - x_k) \, 2 \epsilon.
\label{eq:34}
\end{equation}

Substitution of (\ref{eq:34}) into (\ref{eq:ODE_set}) gives
\begin{equation}
\mathcal{L} \frac{y_k(x)}{ f(x_k) \, 2 \epsilon} = \delta(x - x_k).
\label{eq:35}
\end{equation}

Comparison of (\ref{eq:35}) with (\ref{eq:Green_fn_1D}) makes it clear that
\begin{equation}
G(x, x_k) = \frac{y_k(x)}{ f(x_k) \, 2 \epsilon},
\label{eq:36}
\end{equation}
and that the total solution
\begin{equation}
y(x) = \sum_k y_k(x) = \sum_k G(x, x_k) \, f(x_k) \, 2 \epsilon,
\label{eq:solution_integral}
\end{equation}
in the limit $\epsilon \to 0$, is equal to the integral in (\ref{eq:ODE_solution})

In conclusion, our analysis of the Green function method reveals the fact that we have used of a Dirac delta function
(\ref{eq:Dirac_delta_d}) that obeys the weak definition. Only in this way we were able to split the source term $f(x)$,
which is a continuous function, into a sum of terms, each of them proportional to a Dirac delta function,
as seen in (\ref{eq:34}). 

\section{The Schwarzschild-Tetrode-Fokker action}

Another strong argument in favor of the weak definition of the Dirac delta function is provided by a
discussion of the Schwarzschild-Tetrode-Fokker (STF) action. 

Schwartzschild \cite{Schwarzschild1903} has discovered that the Lorentz force can be obtained 
from the Lagrangian \cite{Goldstein1980}
\begin{equation}
L = \frac{1}{2} m v^2 - q \phi + \frac{q}{c} {\bf A}\cdot{\bf v},
\label{eq:L_1}
\end{equation}
which takes the form \cite{Goldstein1980}
\begin{equation}
L = - m c^2 \sqrt{1 - \frac{v^2}{c^2}} - q \phi + \frac{q}{c} {\bf A}\cdot{\bf v},
\label{eq:L_2}
\end{equation}
in Special Relativity. 
Since $U_\alpha A^\alpha = \gamma \phi c - \gamma {\bf A}\cdot{\bf v}$, 
the product of the Lagrangian (\ref{eq:L_2}) with the Lorentz factor $\gamma$
is a Lorentz invariant \cite{Jackson1999}.
\begin{equation}
L \gamma = - m c^2  - \frac{q}{c} U_\alpha A^\alpha,
\label{eq:L_3}
\end{equation}

Introducing the proper time, and since $dt = \gamma \, d\tau$, the action integral takes the form 
\begin{equation}
\int_{t_i}^{t_f} L \, dt = \int_{\tau_i}^{\tau_f} \Big( - m c^2  - \frac{q}{c} U_\alpha A^\alpha \Big) d\tau.
\label{eq:action1}
\end{equation}

Tetrode \cite{tetrode}, realizing the importance of time symmetric electrodynamics, 
has replaced the retarded LW four-potential 
with the arithmetic mean of the retarded and advanced potentials.
Using the Dirac delta function identity (\ref{eq:Dirac_delta_property_1}), 
from (\ref{eq:Green_fn_solution_2}) and (\ref{eq:Green_fn_solution_2_adv}) we obtain
\begin{multline}
A^\alpha (x) = \frac{1}{2} A_{(ret)}^\alpha (x) + \frac{1}{2} A_{(adv)}^\alpha (x) \\
= \frac{1}{2} q \int \frac{\delta(c t - c T - R)}{R} \, U^\alpha(\tau) \, d\tau
+ \frac{1}{2} q \int \frac{\delta(c t - c T + R)}{R} \, U^\alpha(\tau) \, d\tau \\
= q \int \delta\big( c^2 (t - T)^2 - ({\bf x} - {\bf X})^2 \big) \, U^\alpha(\tau) \, d\tau
= q \int \delta\big( (x - X)^2 \big) \, U^\alpha(\tau) \, d\tau.
\label{eq:timesympot}
\end{multline}

Quite remarkable, working before Dirac introduced his Dirac delta function in 1927, 
Tetrode introduced his own
definition of the Dirac delta function, which in his notation was written as 
$\delta(\sigma^2) = f(\sigma^2)/\sigma^2$.
He also mentioned that only the infinitesimal neighborhood of the lightcone 
(\lq\lq da{\ss} nur die infinitesimale Umgebung des Lichtkegels\rq\rq) with vertex at the field point
contributes to the electromagnetic potentials, in effect giving the weak definition of the Dirac delta function.

Since in (\ref{eq:action1}) $q$, $U_\alpha$, and $\tau$ refer to the test particle at $x$, 
while in (\ref{eq:timesympot}) $q$, $U^\alpha$, and $\tau$ refer to the source particle at $X$, 
in order to substitute (\ref{eq:timesympot}) into (\ref{eq:action1}) we need a change in notation.
Let us label with \lq\lq1\rq\rq\ the test particle, and with \lq\lq2\rq\rq\ the source particle.
The action integral becomes \cite{tetrode}
\begin{equation}
W_1 = - \int_{\tau_{1 i}}^{\tau_{1 f}}  m_1 c^2 \, d\tau_1
 - \frac{q_1 q_2}{c} \int_{\tau_{1 i}}^{\tau_{1 f}} \int 
\delta\big( (x_1 - x_2)^2 \big) \, U_{1 \alpha}(\tau_1) \, U_2^\alpha(\tau_2) \, d\tau_2 \, d\tau_1.
\label{eq:action2}
\end{equation}
Tetrode writes down the negative of (\ref{eq:action2}), which he acknowledges,
but this has no effect on the equations of motion obtained from the extremum condition.
The interaction term in Tetrode's action integral is still negative, due to the fact that
he is using the $(i c t, x, y, z)$ Minkowski formalism, equivalent to a metric tensor of
signature $(-, +, +, +)$, while we, following Jackson \cite{Jackson1999}, 
are using a metric tensor of signature $(+, -, -, -)$.

Tetrode then notices that the interaction term in (\ref{eq:action2}) is symmetric in the indices of the two 
particles. For an infinitesimal variation of $x_1$ 
that vanishes at $\tau_{1 i} = - \infty$ and at $\tau_{1 f} = \infty$, 
the condition $\delta W_1 = 0$ gives the equations of
motion of particle 1 in the electromagnetic field of particle 2. The same holds true for the motion
of particle 2 in the field of particle 1, provided that the whole action integral is also symmetric
in the indices of the particles.
\begin{multline}
W_{12} = - \int  m_1 c^2 \, d\tau_1 - \int  m_2 c^2 \, d\tau_2 \\
 - \frac{q_1 q_2}{c} \int \int 
\delta\big( (x_1 - x_2)^2 \big) \, U_{1 \alpha}(\tau_1) \, U_2^\alpha(\tau_2) \, d\tau_2 \, d\tau_1.
\label{eq:action3}
\end{multline} 

In the final step Tetrode generalizes the action integral (\ref{eq:action3}) to any number of electrically charged interacting particles.
Since $c \, d\tau = \sqrt{dx_\alpha dx^\alpha}$ and $U_\alpha \, d\tau = dx_\alpha$, the action function 
for the whole world becomes
\begin{multline}
W = - \sum_A \int  m_A c \sqrt{dx_{A \alpha} dx_A^\alpha} \\
 - \sum_A \sum_{B > A} \frac{q_A q_B}{c} \int \int 
\delta\big( (x_{A \beta} - x_{B \beta}) (x_A^\beta - x_B^\beta) \big) \, dx_{A \alpha} \, dx_B^\alpha.
\label{eq:action4}
\end{multline}

This is the exact expression from which Wheeler and Feynman started their theory of
time symmetric action-at-a-distance electrodynamics \cite{Wheeler1949}. Because they have used
a metric tensor of signature $(-, +, +, +)$, the scalar products bring in sign changes, and they
warn the readers about this issue in a footnote. In the same footnote they also give the definition of the
Dirac delta function, but this time, unlike Tetrode, they give the strong definition. 
What difference does it make? The strong definition implies that the interaction takes place between
point particles, points in the 4D Minkowski space. The weak definition implies that the interaction
takes place between length elements, infinitesimal segments along the worldlines of the particles.

Fokker \cite{fokker}, deriving his results without the help of the Dirac delta function, completely avoids 
the integrals from (\ref{eq:Green_fn_solution_2}) and (\ref{eq:Green_fn_solution_2_adv}) by
directly inserting into (\ref{eq:action1}) the expression of the time symmetric four-potential,
written with the help of (\ref{eq:LWpotentials}) and (\ref{eq:LWpotentials_adv}) as
\begin{multline}
A^\alpha (x) = \frac{1}{2} A_{(ret)}^\alpha (x) + \frac{1}{2} A_{(adv)}^\alpha (x) \\
= \frac{q}{2 R(T_r)} \frac{\big( 1, {\bf V}(T_r)/c \big)}{1 - \frac{{\bf V}(T_r) \cdot {\bf R}(T_r)}{c \, R(T_r)}}
+ \frac{q}{2 R(T_a)} \frac{\big( 1, {\bf V}(T_a)/c \big)}{1 + \frac{{\bf V}(T_a) \cdot {\bf R}(T_a)}{c \, R(T_a)}} \\
= \frac{q}{2} \frac{U^\alpha(T_r)}{\big( x_\beta - X_\beta(T_r) \big) U^\beta(T_r)}
- \frac{q}{2} \frac{U^\alpha(T_a)}{\big( x_\beta - X_\beta(T_a) \big) U^\beta(T_a)},
\label{eq:timesympot2}
\end{multline}
where $x - X(T_r) = \big( c t - c T_r, {\bf x} - {\bf X}(T_r) \big) = \big( R(T_r), {\bf R}(T_r) \big)$
and $x - X(T_a) = \big( c t - c T_a, {\bf x} - {\bf X}(T_a) \big) = \big( - R(T_a), {\bf R}(T_a) \big)$.
Fokker has used a metric tensor with signature $(+, -, -, -)$ and reduced electrostatic units, 
instead of Gaussian units. This brings an extra $4 \pi$ in the denominator.

After we label with \lq\lq1\rq\rq\ the test particle, and with \lq\lq2\rq\rq\ the source particle,
the action integral becomes
\begin{multline}
W_{12} = - \int  m_1 c^2 \, d\tau_1 - \int  m_2 c^2 \, d\tau_2 \\
 - \frac{q_1 q_2}{2 c} \int \bigg(
\frac{U_{1 \alpha}(\tau_1) \, U_2^\alpha(T_{2 r})}{\big( x_{1 \beta} - x_{2 \beta}(T_{2 r}) \big) U_2^\beta(T_{2 r})}
- \frac{U_{1 \alpha}(\tau_1) \, U_2^\alpha(T_{2 a})}{\big( x_{1 \beta} - x_{2 \beta}(T_{2 a}) \big) U_2^\beta(T_{2 a})}
\bigg) d\tau_1.
\label{eq:action5}
\end{multline}

Based on Tetrode's expression (\ref{eq:action3}), we expect Fokker's expression (\ref{eq:action5}) 
to also be symmetric in the indices of the two particles. A permutation of the two indices gives
\begin{multline}
W_{21} = - \int  m_1 c^2 \, d\tau_1 - \int  m_2 c^2 \, d\tau_2 \\
 - \frac{q_1 q_2}{2 c} \int \bigg(
\frac{U_{2 \alpha}(\tau_2) \, U_1^\alpha(T_{1 r})}{\big( x_{2 \beta} - x_{1 \beta}(T_{1 r}) \big) U_1^\beta(T_{1 r})}
- \frac{U_{2 \alpha}(\tau_2) \, U_1^\alpha(T_{1 a})}{\big( x_{2 \beta} - x_{1 \beta}(T_{1 a}) \big) U_1^\beta(T_{1 a})}
\bigg) d\tau_2.
\label{eq:action6}
\end{multline}

The two action integrals (\ref{eq:action5}) and (\ref{eq:action6}) are indeed the same when 
the retarded part of (\ref{eq:action5}) is equal to the advanced part of (\ref{eq:action6}),
while at the same time the advanced part of (\ref{eq:action5}) is equal to the retarded part of (\ref{eq:action6}).
We must have
\begin{equation}
\frac{d\tau_1}{( x_{1 \beta} - x_{2 \beta} ) U_2^\beta} 
= \frac{d\tau_2}{( x_{1 \beta} - x_{2 \beta} ) U_1^\beta},
\label{eq:segments}
\end{equation}
regardless of whether $t_1 > t_2$ or $t_1 < t_2$. Equivalently, we can write (\ref{eq:segments}) as
\begin{equation}
( x_{1 \beta} - x_{2 \beta} ) \, dx_2^\beta = ( x_{1 \beta} - x_{2 \beta} ) \, dx_1^\beta,
\label{eq:segments2}
\end{equation}
an equation that describes two infinitesimal segments $dx_1$ and $dx_2$ whose endpoints are connected
by light signals. Indeed, this is the equation that emerges when we write
\begin{eqnarray}
( x_{1 \beta} - x_{2 \beta} ) ( x_1^\beta - x_2^\beta ) = 0, \\
( x_{1 \beta} + dx_{1 \beta} - x_{2 \beta} - dx_{2 \beta} ) 
( x_1^\beta + dx_1^\beta - x_2^\beta - dx_2^\beta ) = 0, 
\label{eq:lamont}
\end{eqnarray}
and we keep only the first order terms in the infinitesimals \cite{lamont}.
In Fokker's article \cite{fokker}, with his notation, equation (\ref{eq:segments2}) is written as
$( R \cdot dy ) = ( R \cdot dx )$, and the infinitesimal segments connected by light rays are called
corresponding effective elements (\lq\lq entsprechenden effektiven Elementen\rq\rq).

Multiplying, in the numerator and in the denominator, 
the retarded part of (\ref{eq:action5}) by a retarded $d\tau_{2 r}$ satisfying (\ref{eq:segments}),
and multiplying the advanced part by an advanced $d\tau_{2 a}$ also satisfying (\ref{eq:segments}),
we are able to write down the action function as
\begin{multline}
W_{12} = - \int  m_1 c^2 \, d\tau_1 - \int  m_2 c^2 \, d\tau_2 \\
 - \frac{q_1 q_2}{2 c} \int \bigg(
\frac{dx_{1 \alpha} \, dx_{2 r}^\alpha}{\big( x_{1 \beta} - x_{2 r \beta} \big) dx_{2 r}^\beta}
- \frac{dx_{1 \alpha} \, dx_{2 a}^\alpha}{\big( x_{1 \beta} - x_{2 a \beta} \big) dx_{2 a}^\beta}
\bigg),
\label{eq:action7}
\end{multline}
in a way that stresses the fact that 
the electromagnetic interaction takes place between corresponding infinitesimal segments.
In the final step Fokker generalizes the action integral (\ref{eq:action7}) 
to any number of electrically charged interacting particles.

Dirac himself recognized that the action integral admits two equivalent formulations, 
writing it without the Dirac delta function \cite{dirac1938}, like Fokker,   
and with the Dirac delta function \cite{dirac1942}, like Tetrode.
In this way the STF action,
describing the interaction between corresponding infinitesimal segments,
reveals the fact that the Dirac delta function used in electrodynamics is the one that obeys the weak definition.

As a side note, Dirac \cite{dirac1942} also recognized that, in this context, the use of a Dirac delta function
obeying the strong definition is problematic, and suggested replacing it,
in the STF action, with a different function $\delta^*$
that, just like the Dirac delta function obeying the weak definition, depends on an 
infinitesimal parameter ${\boldsymbol \lambda}$ that has the limit ${\boldsymbol \lambda} \to 0$.

\section{Concluding remarks}

In summary, the conclusion that we have reached based on our 
algebraic derivation of the Doppler factor in the LW potentials of a point charged particle
is in full agreement with the conclusion that we have reached based on our geometrical derivation \cite{Galeriu2021a}.
The point particles are points only in the 3D Euclidean space, while in the 4D Minkowski space the 
electromagnetic interaction takes place between corresponding infinitesimal segments along the worldlines
of the particles, just as Fokker has described in his groundbreaking article \cite{fokker}. 
We have discovered that the Dirac delta function used in electrodynamics does not
obey the strong definition, as it is generally assumed, but instead it obeys the weak definition.
This conclusion emerges from our analysis of the Dirac delta function identity
responsible for the apparition of the Doppler factor,
from our analysis of the Green function method used in the derivation of the LW potentials, 
and from our analysis of the STF action.

This is not the first time when, in electrodynamics, the use
of the Dirac delta function (\ref{eq:Dirac_delta_a}) that is non-zero in just one point has been questioned. 
As early as 1935 Louis de Broglie \cite{deBroglie1935} has suggested replacing this
infinitely narrow needle-like function (\lq\lq une fonction en aiguille infiniment fine\rq\rq) 
by a needle-like function of very small but finite width (\lq\lq une fonction en aiguille d'\'{e}paisseur tr\`{e}s petite, mais finie\rq\rq),
like for example
\begin{equation}
\delta(w) = \frac{1}{\sigma \sqrt{\pi}} e^{- \frac{w^2}{\sigma^2}},
\label{eq:debroglie}
\end{equation}
where $\sigma$ is a very small and positive parameter. This function is continuous and differentiable,
and can be easily generalized to higher dimensions, producing a Lorentz invariant expression.
The same idea, as explained in detail by Feynman \cite{FeynmanII}, has been explored by Fritz Bopp.
Feynman also encourages us to study such theories, in order \lq\lq to see the struggles of the human mind\rq\rq.
We hope that at least this goal was achieved!

\section*{Acknowledgments}

The author is very much indebted to David~H.~Delphenich for 
translating the articles written by Tetrode and Fokker, upon personal request.

\end{document}